\documentclass[aps, prstab, reprint, floatfix,
  endfloats*,
  preprintnumbers, showpacs, showkeys]{revtex4-1}
\usepackage{graphicx}
\usepackage[colorlinks=true, linkcolor=blue, citecolor=blue,
filecolor=blue, urlcolor=blue]{hyperref}
\usepackage{mathptmx}

\newcommand{\q}[2]{\ensuremath{#1\ \mathrm{#2}}}

\newcommand{\Lu}{\ensuremath{{\cal L}}}

\newcommand{\Np}{\ensuremath{N^p}}
\newcommand{\Na}{\ensuremath{N^a}}

\newcommand{\epx}{\ensuremath{\varepsilon^p_x}}
\newcommand{\epy}{\ensuremath{\varepsilon^p_y}}
\newcommand{\eax}{\ensuremath{\varepsilon^a_x}}
\newcommand{\eay}{\ensuremath{\varepsilon^a_y}}
\newcommand{\epz}{\ensuremath{\varepsilon^p_z}}
\newcommand{\eaz}{\ensuremath{\varepsilon^a_z}}

\newcommand{\qp}{\ensuremath{Q^p}}
\newcommand{\qa}{\ensuremath{Q^a}}
\newcommand{\qpx}{\ensuremath{Q^p_x}}
\newcommand{\qpy}{\ensuremath{Q^p_y}}
\newcommand{\qax}{\ensuremath{Q^a_x}}
\newcommand{\qay}{\ensuremath{Q^a_y}}

\newcommand{\xip}{\ensuremath{\xi^p}}
\newcommand{\xia}{\ensuremath{\xi^a}}
\newcommand{\xipx}{\ensuremath{\xi^p_x}}
\newcommand{\xipy}{\ensuremath{\xi^p_y}}
\newcommand{\xiax}{\ensuremath{\xi^a_x}}
\newcommand{\xiay}{\ensuremath{\xi^a_y}}

\begin{document}

\date{March 14, 2012}

\title{Bunch-by-bunch measurement of transverse coherent beam-beam modes\\
in the Fermilab Tevatron collider}

\author{Giulio~Stancari}
\thanks{Corresponding author}
\email[e-mail: ]{stancari@fnal.gov}
\altaffiliation{on leave from Istituto Nazionale di Fisica Nucleare
  (INFN), Sezione di Ferrara, Italy.}
\author{Alexander~Valishev}
\affiliation{Fermi National Accelerator Laboratory, P.O. Box 500,
Batavia, IL 60510, U.S.A.}

\begin{abstract}
  A system for bunch-by-bunch detection of transverse proton and
  antiproton coherent oscillations in the Tevatron is described. It is
  based on the signal from a single beam-position monitor located in a
  region of the ring with large amplitude functions. The signal is
  digitized over a large number of turns and Fourier-analyzed offline
  with a dedicated algorithm. To enhance the signal, band-limited
  noise is applied to the beam for about 1~s. This excitation does not
  adversely affect the circulating beams even at high
  luminosities. The device has a response time of a few seconds, a
  frequency resolution of $1.6\times 10^{-5}$ in fractional tune, and
  it is sensitive to oscillation amplitudes of 60~nm. It complements
  Schottky detectors as a diagnostic tool for tunes, tune spreads, and
  beam-beam effects. Measurements of coherent mode spectra are
  presented to show the effects of betatron tunes, beam-beam
  parameter, and collision pattern, and to provide an experimental
  basis for beam-beam numerical codes. Comparisons with a simplified
  model of beam-beam oscillations are also described.
\end{abstract}



\preprint{\emph{FERMILAB-PUB-11-181-APC, accepted for publication in
  Phys.\ Rev.\ ST Accel.\ Beams (April 2012 issue)}}

\maketitle

\section{Introduction}

In particle colliders, each beam experiences nonlinear forces when
colliding with the opposing beam. A manifestation of these forces is a
vibration of the bunch centroids around the closed orbit.  These
coherent beam-beam oscillation modes were observed in several lepton
machines, including PETRA, TRISTAN, LEP, and
VEPP-2M~\cite{Piwinski:IEEE:1979, Ieiri:NIM:1988, Keil:ICHEA:1992,
  Nesterenko:PRE:2002}. Although their observation in hadron machines
is made more challenging by the lack of strong damping mechanisms to
counter external excitations, they were seen both at the ISR and at
RHIC~\cite{Koutchouk:CERN:1982a, Koutchouk:CERN:1982b,
  Fischer:BNL:2002, Fischer:PAC:2003, Pieloni:PhD:2008}.  Originally,
one motivation for the study of coherent beam-beam modes was the
realization that their frequencies may lie outside the incoherent tune
distribution, with a consequent loss of Landau
damping~\cite{Alexahin:PA:1999}. The goal of the present research is
to develop a new diagnostic tool to estimate bunch-by-bunch tune
distributions, to assess the effects of Gaussian electron lenses for
beam-beam compensation~\cite{Shiltsev:PRSTAB:1999,Shiltsev:NJP:2008,
  Shiltsev:PRSTAB:2008, Valishev:PAC:2011}, and to provide an
experimental basis for the development of beam-beam numerical codes.

The behavior of colliding bunches is analogous to that of a system of
oscillators coupled by the beam-beam force. In the simplest case, when
2~identical bunches collide head-on in one interaction region,
2~normal modes appear: a $\sigma$~mode (or 0~mode) at the lattice
tune, in which bunches oscillate transversely in phase, and a
$\pi$~mode, separated from the $\sigma$~mode by a shift of the order
of the beam-beam parameter, in which bunches are out of phase. In
general, the number, frequency and amplitude of these modes depend on
the number of bunches, on the collision pattern, on the tune
separation between the two beams, on transverse beam sizes and on
relative intensities. Coherent beam-beam modes have been studied at
several levels of refinement, from analytical linear models to fully
3-dimensional particle-in-cell calculations~\cite{Piwinski:IEEE:1979,
  Meller:IEEE:1981, Yokoya:PA:1990, Alexahin:NIM:2002,
  Herr:PRSTAB:2001, Pieloni:PAC:2005, Qiang:NIM:2006,
  Pieloni:PhD:2008, Stern:PRSTAB:2010}.

In the Tevatron, 36 proton bunches (identified as P1--P36) collide
with 36 antiproton bunches (A1--A36) at the center-of-momentum energy
of 1.96~TeV. There are 2 head-on interaction points (IPs),
corresponding to the CDF and the DZero experiments. Each particle
species is arranged in 3~trains of 12~bunches each, circulating at a
revolution frequency of 47.7~kHz. The bunch spacing within a train is
396~ns, or 21 53-MHz rf buckets. The bunch trains are separated by
2.6-$\mu$s abort gaps. The synchrotron frequency is 34~Hz, or $7\times
10^{-4}$ times the revolution frequency. The machine operates with
betatron tunes near~20.58.

The betatron tunes and tune spreads of individual bunches are among
the main factors that determine beam lifetimes and collider
performance. They are affected by head-on and long-range beam-beam
interactions. Three systems are currently used in the Tevatron to
measure incoherent tune distributions: the 21.4-MHz Schottky
detectors, the 1.7-GHz Schottky detectors, and the direct diode
detection base band tune (3D-BBQ). The latter two can be gated on
single bunches. Detection of transverse coherent modes can complement
these three systems because of its sensitivity, bunch-by-bunch
capability, high frequency resolution, and fast measurement time.

The basis for the measurement technique was presented in
Ref.~\cite{Carneiro:Beamsdoc:2005}, and preliminary results can be
found in
Refs.~\cite{Semenov:PAC:2007,Kamerdzhiev:BIW:2008,Valishev:EPAC:2008}.
Several improvements, mainly in the data analysis, were implemented
and presented in a concise report~\cite{Stancari:BIW:2010}. In this
paper, we describe the detection technique in detail. We also present
a wide set of measurements illustrating the performance of the device
and the response of the coherent mode spectra to various experimental
conditions, such as betatron tune separation, beam-beam parameter, and
collision pattern.

\section{Modeling}
\label{sec:model}

\begin{figure}
\begin{tabular}[c]{cc}
  \parbox{4mm}{\rotatebox{90}{Fractional tune}} &
  \parbox{78mm}{\includegraphics[width=78mm]{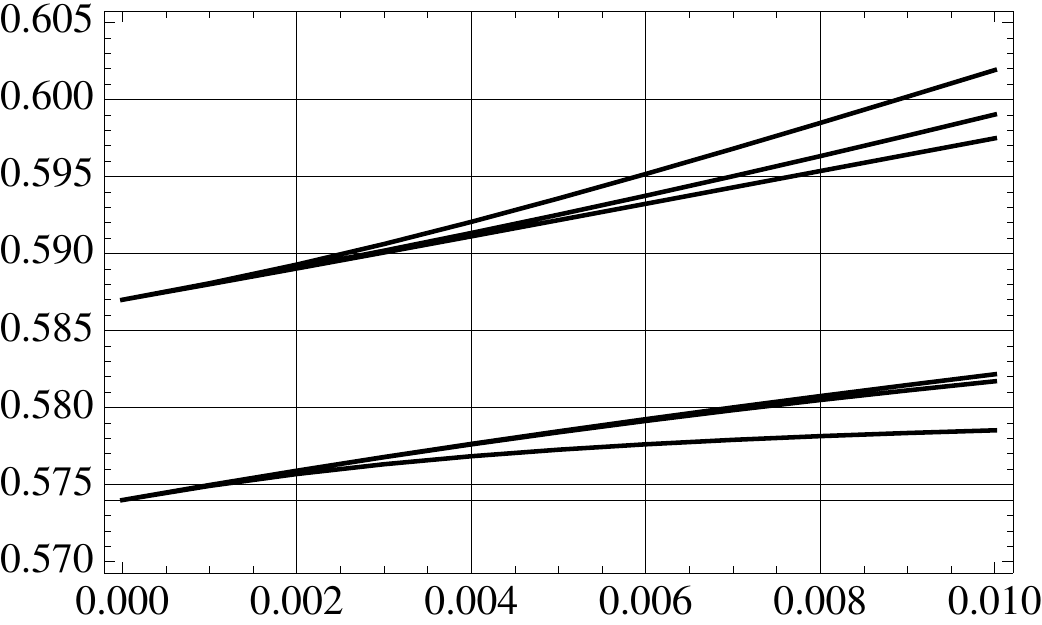}} \\
  & Beam-beam parameter, $\xi$
\end{tabular}
\caption{Coherent mode tunes vs. beam-beam parameter calculated with the
linearized model; $\qp = 0.587$, $\qa =0.574$, $\xi = \xip = \xia$.}
\label{fig:MatrixTunes}
\end{figure}

The basic features of transverse coherent oscillations can be
described by a simple model. In the Tevatron, these oscillations are
substantially nonlinear due to the properties of the lattice and of
the beam-beam force. Hence, the rigid bunch approximation cannot
provide an accurate view of the coherent mode spectrum. However, this
approximation can be used for qualitative analysis of the expected
beam-beam mode tunes and their dependence on the the betatron
tunes~$Q$ and the beam-beam parameter per interaction point~$\xi$.

We use a simple matrix formalism to compute the eigenmode tunes of the
system of colliding bunches. Besides employing the rigid bunch
approximation, one more simplification is used. The complete
description of the system would require modeling the interaction of
72~bunches at 138 collision points. The analysis of such a system can
be quite complex. Observations and analytical estimates show that the
difference in tunes between individual bunches is small compared to
the beam-beam tune shift. Thus, as a first approximation, it is
possible to neglect long range interactions.  This limits the system
to 6~bunches (3 in each beam) colliding at two head-on interaction
points. In the following discussion, we limit betatron oscillations to
one degree of freedom. Because the system has 3-fold symmetry, the
1-turn map transporting the 12-vector of dipole moments and momenta of
the system of 6~bunches can be expressed as follows:
\begin{equation}
M = M_\mathrm{BB3} \  M_\mathrm{T3} \  M_\mathrm{BB2} \  M_\mathrm{T2}
\  M_\mathrm{BB1} \  M_\mathrm{T1} ,
\end{equation}
where $M_\mathrm{TN}$ ($N =1,2,3$) are the 2$\times$2 block-diagonal
12$\times$12 matrices transporting phase space coordinates through the
accelerator arcs, and $M_\mathrm{BBN}$ are the matrices describing
thin beam-beam kicks at the IPs. Although there are only
2~interactions per bunch, 3~collision matrices are used to describe a
one-turn map of the system of six bunches. This construction
represents the time propagation of the bunch coordinates through one
turn with break points at the CDF (B0), D0 and F0 locations in the
machine. If on a given step the bunch is at B0 or D0, its momentum
coordinate is kicked according to the distance between the centroids
of this bunch and of the opposing bunch. If the bunch is at F0 (1/3 of
the circumference from B0 and D0), where the beams are separated, its
momentum is unchanged. For example, the matrix describing the
interaction of proton bunch~1 with antiproton bunch~2 at CDF and
proton bunch~3 with antiproton bunch~3 at DZero has the following
form:

\begin{widetext}
\begin{equation}
M_\mathrm{BB1} = \left(
\begin{array}{cccccccccccc}
  1 & 0 & 0 & 0 & 0 & 0 & 0 & 0 & 0 & 0 & 0 & 0 \\
  -2\pi\xip/\beta & 1 & 0 & 0 & 0 & 0 & 0 & 0 & 2\pi\xip/\beta & 0 & 0 & 0 \\
  0 & 0 & 1 & 0 & 0 & 0 & 0 & 0 & 0 & 0 & 0 & 0 \\
  0 & 0 & 0 & 1 & 0 & 0 & 0 & 0 & 0 & 0 & 0 & 0 \\
  0 & 0 & 0 & 0 & 1 & 0 & 0 & 0 & 0 & 0 & 0 & 0 \\
  0 & 0 & 0 & 0 & -2\pi\xip/\beta & 1 & 0 & 0 & 0 & 0 & 2\pi\xip/\beta & 0 \\
  0 & 0 & 0 & 0 & 0 & 0 & 1 & 0 & 0 & 0 & 0 & 0 \\
  0 & 0 & 0 & 0 & 0 & 0 & 0 & 1 & 0 & 0 & 0 & 0 \\
  0 & 0 & 0 & 0 & 0 & 0 & 0 & 0 & 1 & 0 & 0 & 0 \\
  2\pi\xia/\beta & 0 & 0 & 0 & 0 & 0 & 0 & 0 & -2\pi\xia/\beta & 1 & 0 & 0 \\
  0 & 0 & 0 & 0 & 0 & 0 & 0 & 0 & 0 & 0 & 1 & 0 \\
  0 & 0 & 0 & 0 & 2\pi\xia/\beta & 0 & 0 & 0 & 0 & 0 & -2\pi\xia/\beta & 1
\end{array}
\right).
\end{equation}
\end{widetext}

Here, \xip\ and \xia\ are the beam-beam parameters for protons and
antiprotons, and $\beta$ is the amplitude function at the IP. The
Yokoya factor~\cite{Yokoya:PA:1990,Yokoya:PRSTAB:2000} is considered
to be equal to~1.  The eigentunes of the 1-turn map are then computed
numerically.

This model provides a quick estimate of the expected values of the
coherent beam-beam mode tunes for a given set of machine and beam
parameters. The model cannot be used for accurate calculation of the
relative amplitude of these modes, which is determined by nonlinear
effects such as Landau damping. For the case of weak nonlinearities,
this approach allows one to determine the mode amplitudes by computing
the projection of mode eigenvectors on the excitation
vector~\cite{Nesterenko:PRE:2002}. In the case of the Tevatron
experiments described below, this is not straightforward because a
wideband noise source was used to excite the beam motion.

In Figure~\ref{fig:MatrixTunes}, an example of the dependence of the
6~eigenfrequencies on the beam-beam parameter per IP is presented. As
one would expect, at small values of $\xi$ (uncoupled oscillators) the
mode frequencies approach the bare lattice tunes; in this case, 0.587
for protons and 0.574 for antiprotons. When the total beam-beam
parameter exceeds the difference between the lattice tunes, the modes
are split and their symmetry approaches that of the conventional
$\sigma$ and $\pi$~modes.  The parameters of this calculation are
taken to resemble those of the beginning of the Tevatron Store~7754,
when the beam-beam parameter was $\xi = \xia = \xip = 0.01$. A
comparison with data is given in Section~\ref{sec:results}
(Figure~\ref{fig:store7754_evolution}).

\section{Apparatus}

\begin{figure}
\includegraphics[width=86mm]{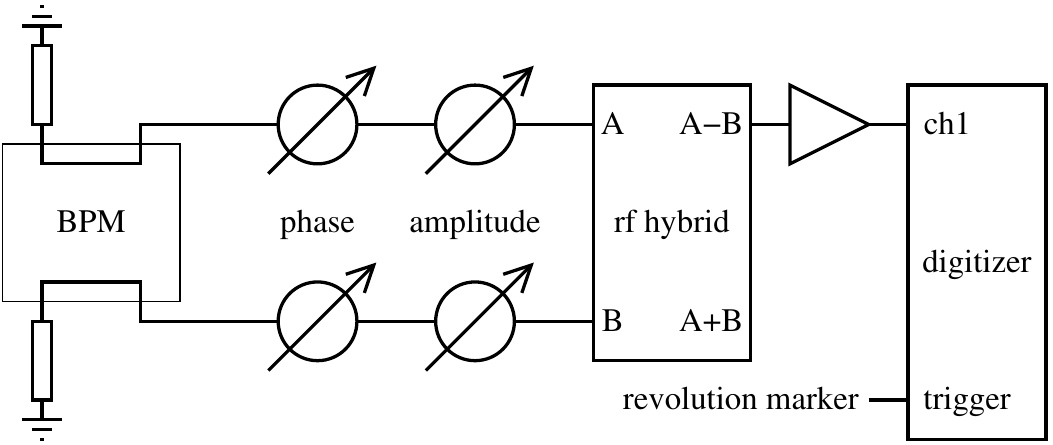}
\caption{Schematic diagram of the apparatus.}
\label{fig:apparatus}
\end{figure}

\begin{figure}
\includegraphics[width=72mm]{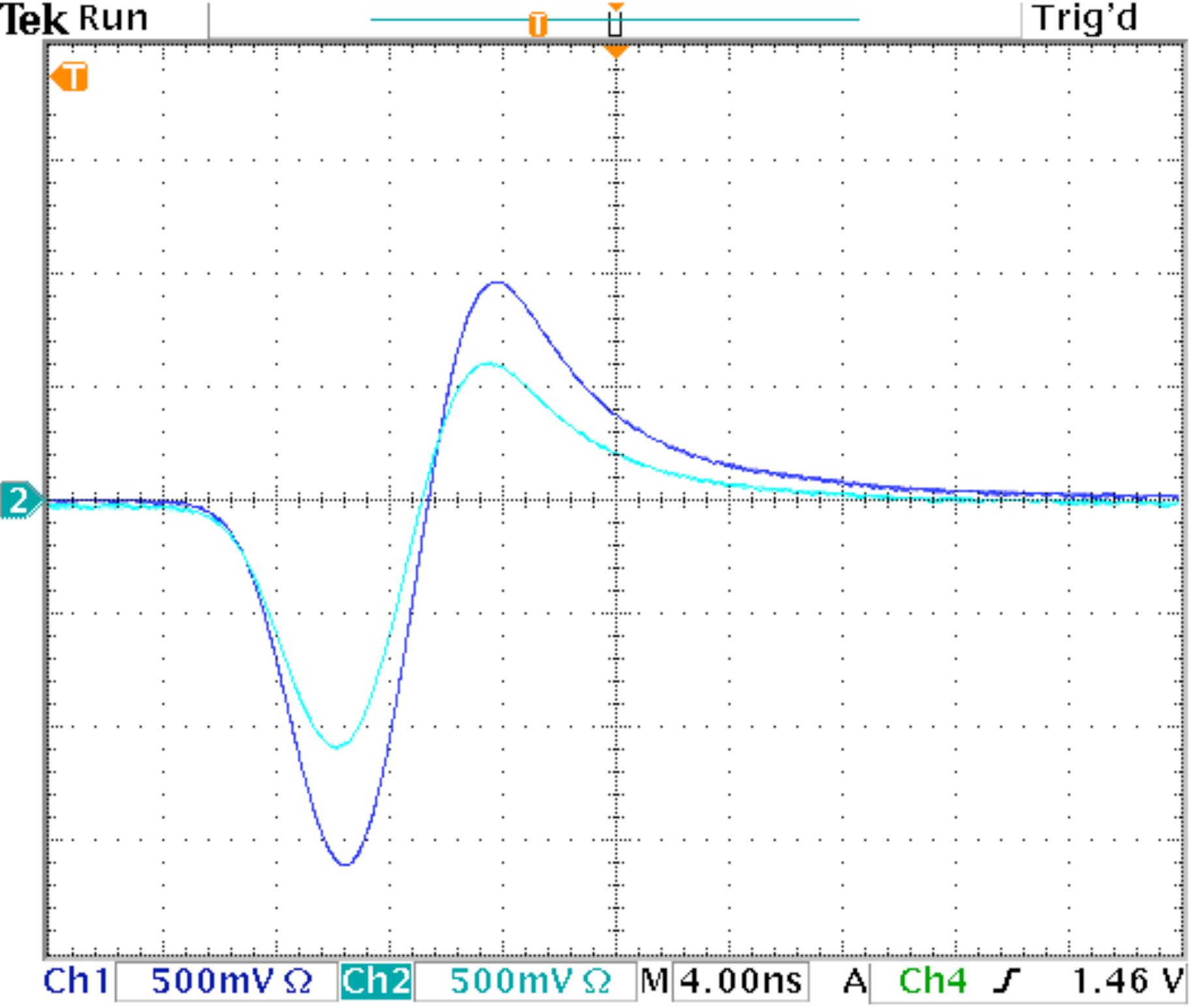} \\
\includegraphics[width=72mm]{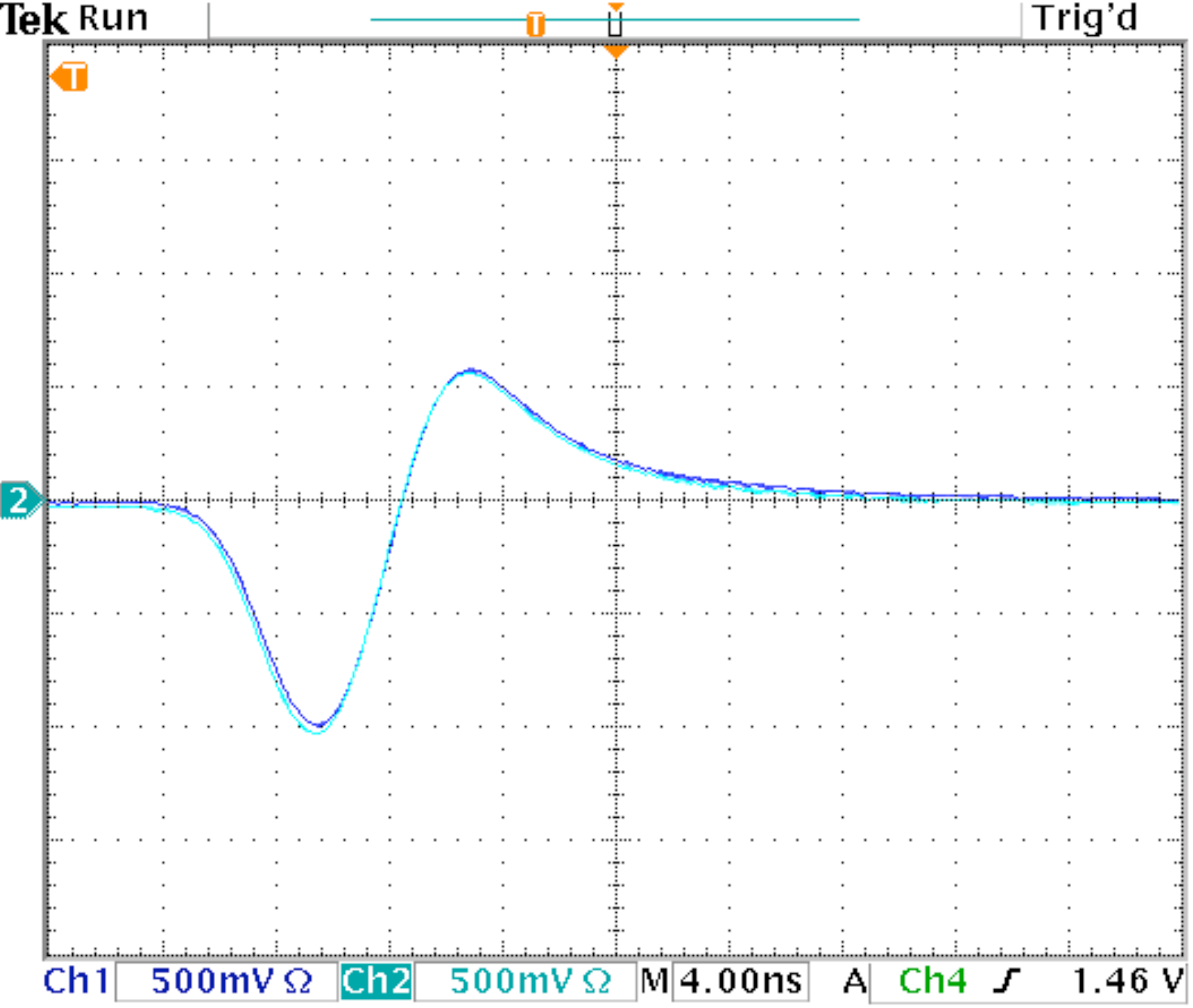} \\
\includegraphics[width=72mm]{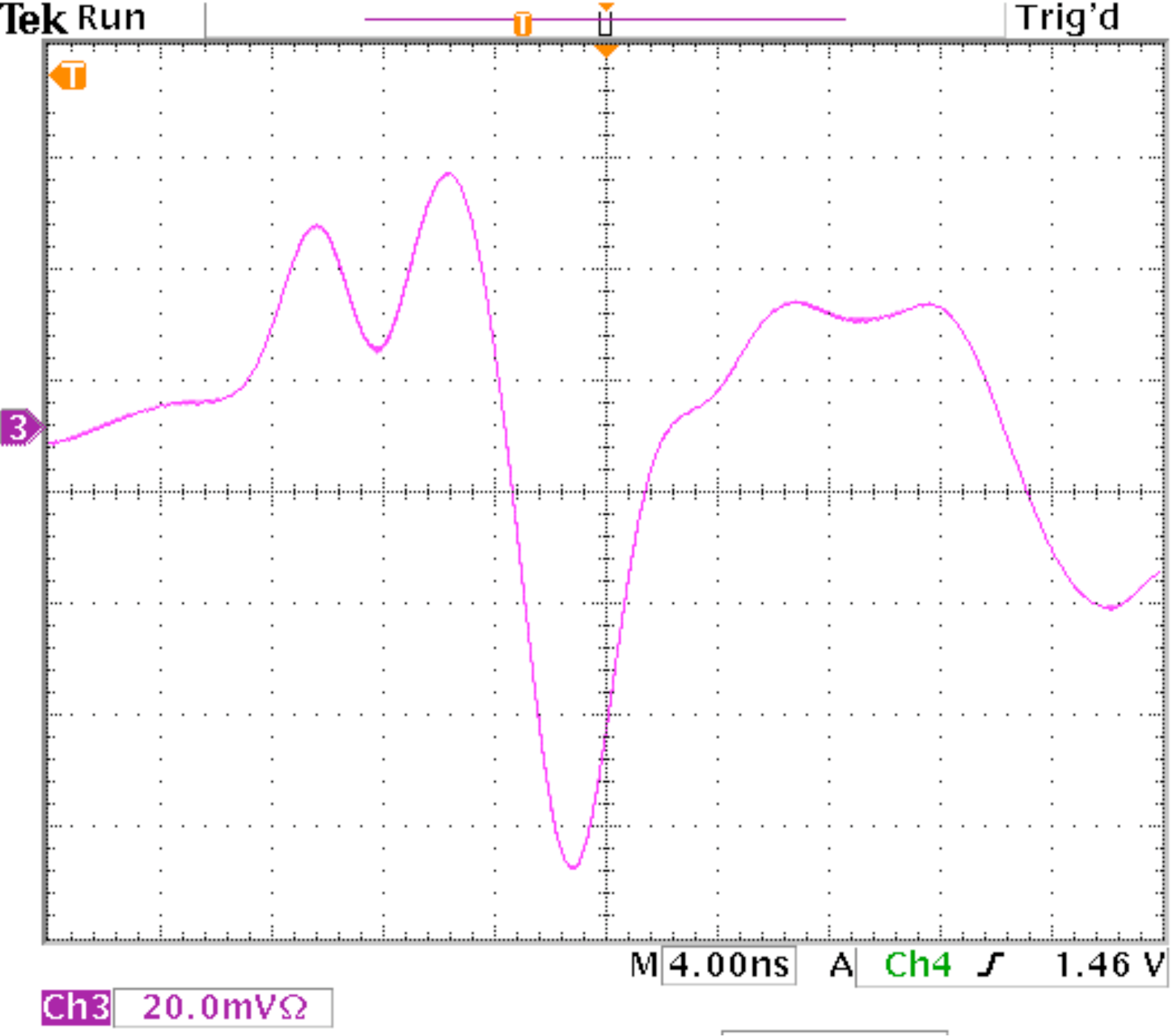}
\caption{BPM signals $A$ (blue) and $B$ (cyan) for an antiproton bunch
  (top); same signals after equalization
  (center); $A-B$ output of the hybrid circuit after amplification
  (magenta, bottom).}
\label{fig:BPM_signals}
\end{figure}

The system for the detection of transverse coherent modes
(Figure~\ref{fig:apparatus}) is based on the signal from a single
vertical beam-position monitor (BPM) located near the CDF interaction
point, in a region where the vertical amplitude function at collisions
is $\beta_y = \q{880}{m}$. The BPM is a stripline pickup, with two
plate outputs ($A$ and $B$) for each of the two counterpropagating
beams.  The proton outputs are split: half of the signal is sent to
the Tevatron BPM readout and orbit stabilization circuits; the other
half is used by the present system. Antiproton signals are about a
factor three weaker and are usually not used for orbit feedback, so
the splitter is not necessary and the full signal can be analyzed.
Switching between proton and antiproton signals presently requires
physically swapping cables.

In the Tevatron, protons and antiprotons share a common vacuum
pipe. Outside of the interaction regions, their orbits wrap around
each other in a helical arrangement. Therefore, bunch centroids can be
several millimeters away from the BPM's electrical axis. Typically,
the peak-to-peak amplitude of the proton signal is 10~V on one plate
and 5~V on the other, whereas the signal of interest is of the order
of a few millivolts.  For this reason, it is necessary to equalize the
$A$ and $B$ signals to take advantage of the full dynamic range of the
digitizer. Equalization also reduces false transverse signals due to
trigger jitter, as discussed below. The phase and attenuation of each
signal is manually adjusted by minimizing the $A-B$ output of the rf
hybrid circuit. If necessary, fine-tuning is done by displacing the
beam with a small orbit bump.  Figure~\ref{fig:BPM_signals} shows an
example of $A$ and $B$ signals after equalization and the $A-B$ output
of the hybrid. Orbits at collisions are stable over a time scale of
weeks, and this manual adjustment does not need to be repeated
often. To automate the task in the case of changing orbits and
intensities (e.g., for observations at top energy between the low-beta
squeeze and initiating collisions, or for observing both proton and
antiproton bunches), a circuit board is being designed with
self-calibrating gains and offsets.

The difference signal from the hybrid is amplified by 23~dB and sent
to the digitizer. We use a 1-channel, 1-V full range, 10-bit digitizer
(Agilent Acqiris series) with time-interleaved analog-to-digital
converters (ADCs). It can sample at 8~GS/s and store a maximum of
1024~MS or 125,000 segments. (Due to a firmware problem, only half of
the segments were used in the experiments described below.) The
47.7-kHz Tevatron revolution marker is used as trigger, so we will
refer to `segments' or `turns' interchangeably. Typically, we sample
at 8~GS/s (sample period of 125~ps), which corresponds to 150~slices
for each 19-ns rf bucket. At this sampling rate, one can record
waveforms of 1~bunch for 62,500 turns, 2~bunches for 52,707 turns, or
12~bunches for 12,382 turns, depending on the measurement of
interest. A C++ program running on the front-end computer controls the
digitizer settings, including its delay with respect to the Tevatron
revolution marker.

Data is written in binary format. The output contains the raw ADC data
together with the trigger time stamps and the delay of the first
sample with respect to the trigger. Timing information has an accuracy
of about 15~ps, and it is extremely important for the synchronization
of samples from different turns.

To enhance the signal, the beam is excited with a few watts of
band-limited noise (`tickling') for about 1~s during the measurement.
The measurement cycle consists of digitizer setup, tickler turn-on,
acquisition start, tickler turn-off, and acquisition stop. The cycle
takes a few seconds. The procedure is parasitical and it does not
adversely affect the circulating beams, even at the beginning of
regular collider stores, with luminosities around \q{3.5\times
  10^{32}}{events/(cm^2\, s)}. When repeating the procedure several
times, the Schottky monitors may show some activity, but no beam loss
is observed.

\section{Data analysis}

\begin{figure*}
\includegraphics[width=0.9\textwidth]{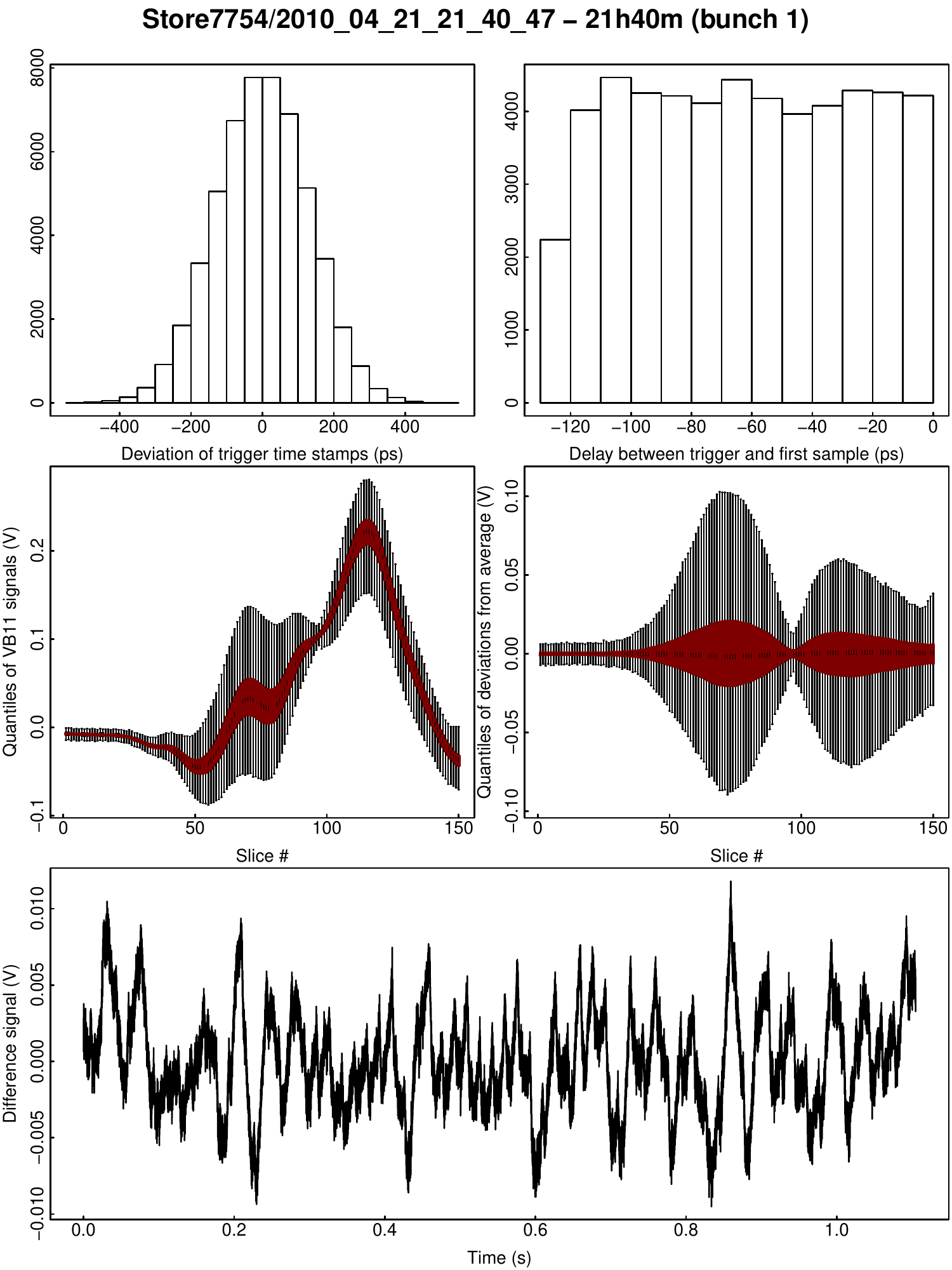}
\caption{Summary plots for one sample data set (Store~7754 at 21:41): difference between recorded trigger time and
  nominal revolution time (top left); recorded offset between trigger
  time stamp and first sample (top right); quantiles (minimum,
  25\%--75\% in red, maximum) of digitized signals over all 52,707
  turns, for each slice (center left); quantiles of digitized signals
  after subtracting each slice's average; average difference signal
  for the signal slices (41--95 and 99--147) over the course of the
  measurement.}
\label{fig:data_summary}
\end{figure*}

\begin{figure*}
\includegraphics[width=\textwidth]{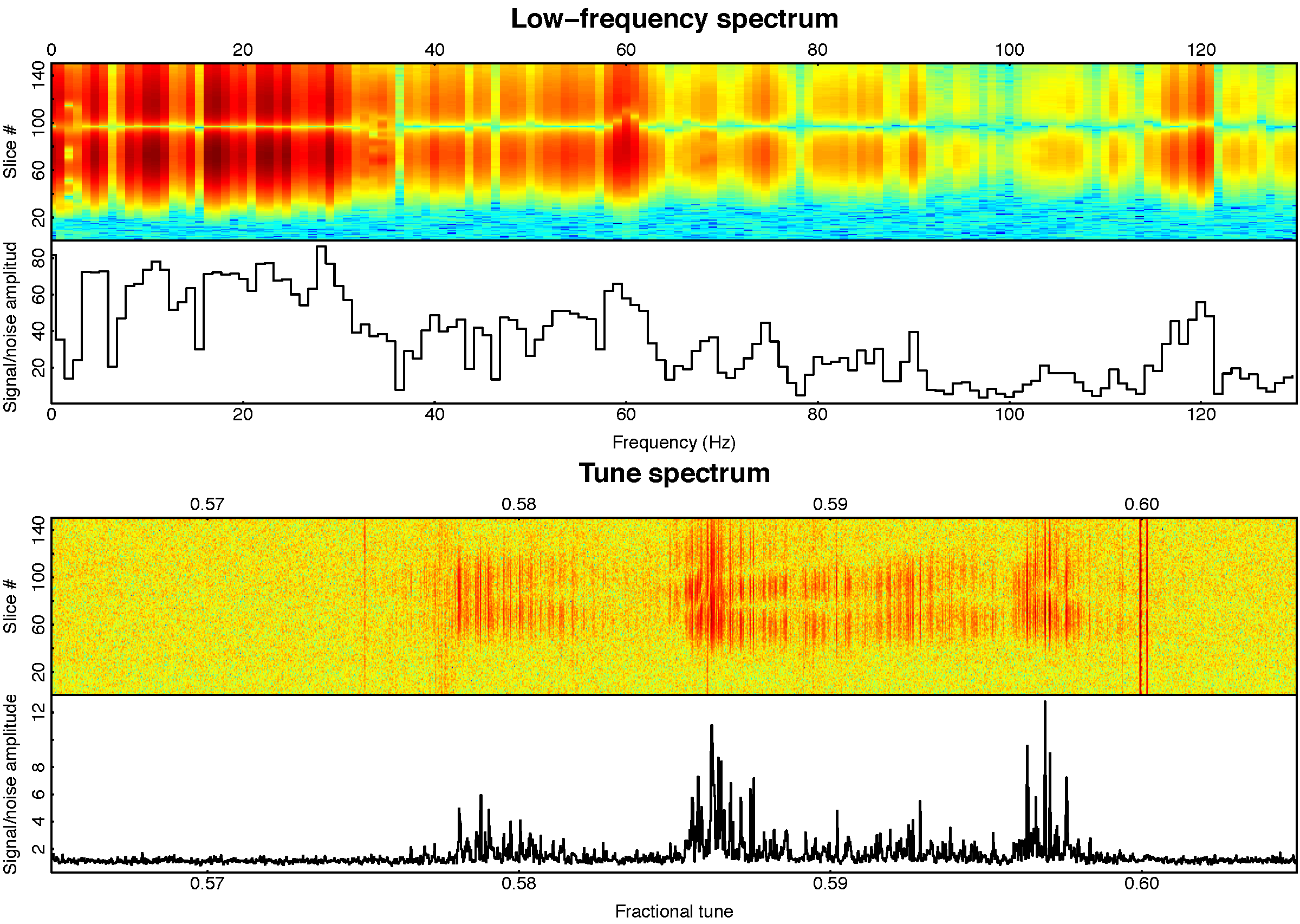}
\caption{Example of frequency spectra for antiprotons from data taken
  during Store~7754. Two selected regions of the spectrum are shown:
  below 130~Hz (top two plots) and around $\q{47.7}{kHz} \times
  (1-0.585) = \q{20}{kHz}$ (bottom two plots). The color plots
  represent the Fourier amplitude (in logarithmic scale) vs.\
  frequency for each slice. The black traces are the average
  amplitudes of the signal slices divided by those of the background
  slices (described in the text).}
\label{fig:analysis_example}
\end{figure*}

Data is analyzed offline using the multi-platform, open-source
R~statistical package~\cite{R:2010}. The distribution of differences
between trigger time stamps from consecutive turns yields the average
revolution frequency (47713.11~Hz at 980~GeV). From it, the nominal or
`ideal' trigger time stamps for each turn are calculated. The
distribution of trigger offsets, i.e. the differences between measured
and nominal time stamps, is a measure of the jitter in the revolution
marker (Figure~\ref{fig:data_summary}, top left). The root mean square
of the distribution is usually less than 0.2~ns.  The delay between
trigger time and the time stamp of the first sample is also recorded
with an accuracy of 15~ps. An offset distribution is shown in
Figure~\ref{fig:data_summary} (top right). As expected for
uncorrelated revolution period and sampling clock, the offsets do not
exceed the sampling period (125~ps, in this case) and their
distribution is flat.  The sum of trigger offset and first-sample
delay is the correction by which each sample in a segment is to be
shifted in time to be aligned with the other segments. For each turn
and each bunch, the signal is interpolated with a natural spline and
shifted in time according to this correction. One undesirable effect
of this synchronization algorithm is that a few slices (usually not
more than 3) at each edge of the bucket become unusable, as they
cannot be replaced with real data. The synchronization of turns is
extremely important, as the jitter in trigger time translates into a
false transverse oscillation where the difference signal has a
slope. If the BPM plates are not perfectly balanced, jitter of even a
fraction of a nanosecond can raise the noise floor by several decibels
and compromise the measurement.

Figure~\ref{fig:data_summary} shows the distribution of digitized data
for each slice in an antiproton bunch after synchronization (center
left) and after subtracting the average for each slice (center
right). Each slice corresponds to 125~ps. At the bottom of
Figure~\ref{fig:data_summary} is the difference signal (proportional
to beam position) over the course of a measurement (52,707 turns, in
this case). Bunch oscillations are dominated by low-frequency beam
jitter attributable to mechanical
vibrations~\cite{Baklakov:PAC:1999}. The range of amplitudes is
inferred from comparisons with the regular Tevatron BPM system and
corresponds to about \q{\pm25}{\mu m}. This low-frequency jitter does
not affect the measurements of coherent beam-beam modes directly, but
it reduces the available dynamic range. A high-pass filter and more
amplification may be employed to improve the system.

For each bunch, the signal of each individual slice vs.\ turn number
is Fourier transformed. Frequency resolution is determined by the
number of bins in the fast Fourier transform (FFT) vector and it is
limited to 62,500 turns, corresponding to $1.6\times 10^{-5}$ of the
revolution frequency or 0.8~Hz. The data is multiplied by a Slepian
window of rank~2 to confine leakage to adjacent frequency bins and
suppress it below $10^{-5}$ in farther bins~\cite{Press:NR:2007}. When
the full frequency resolution is not needed, the FFT vectors are
overlapped by about 1/3 of their length to reduce data loss from
windowing, and the resulting spectral amplitudes are averaged.

Calculations take about 20~s per bunch for 62,500~turns and 150~slices
per bunch on a standard laptop computer (Apple MacBook running Mac OS
X 10.5.8 with 2.4-GHz Intel Core 2 Duo processor and 4~GB of
RAM). Processing time is dominated by the synchronization algorithm.

The noise level is estimated by observing the spectra without
beam. The spectra show a few sharp lines in all slices. These lines
are attributed to gain and offset differences between the
time-interleaved ADCs themselves and to timing skew of their
clocks~\cite{Kurosawa:IEEE:2000, Elbornsson:IEEE:2005,
  Fong:IMTC:2005}. The same spurious lines are also present in the
Fourier spectrum of the time stamps, and this corroborates their
attribution to digitizer noise. To improve the signal-to-noise ratio,
and to suppress backgrounds unrelated to the beam such as the spurious
lines from the digitizer, a set of signal slices (near the signal
peaks) and a set of background slices (before the arrival of the
bunch) are defined. Amplitude spectra are computed for both signal and
background slice sets, and their ratio is calculated. The ratios are
very clean, with some additional variance at the frequencies
corresponding to the narrow noise spikes. Results are presented in
terms of these signal-to-background amplitude ratios.

Figure~\ref{fig:analysis_example} shows an example of analyzed
antiproton data, in two regions of the frequency spectrum: a
low-frequency region with the horizontal axis expressed in hertz (top
two plots) and a high-frequency region, in terms of the revolution
frequency or fractional tune. The 2-dimensional color plots show the
amplitude distribution for each of the 150 125-ps slices in
logarithmic scale. In this example, the signal slices are numbers
41--95 and 99--147. They are defined as the ones for which the
amplitude is above 10\% of the range of amplitudes (see also
Figure~\ref{fig:data_summary}, center right). Background slices are
numbers 3--31 (amplitude below 2\% of range). The black-and-white
1-dimensional plots show the ratio between signal and background
amplitudes.  In the top plots of Figure~\ref{fig:analysis_example},
one can appreciate the strength of the low-frequency components. The
60-Hz power-line noise and its harmonics are also visible. The lines
around 34~Hz and 68~Hz are due to synchrotron oscillations leaking
into the transverse spectrum.  The bottom plots of
Figure~\ref{fig:analysis_example} show the spectra of transverse
coherent oscillations. The vertical lines present in all slices in the
2-dimensinal plot, attributed to digitizer noise, are eliminated by
taking the ratio between signal and background slices. One can also
notice the small variance of the noise level compared to the amplitude
of the signal peaks.

In the 2-dimensional plots of Figure~\ref{fig:analysis_example}, one
may notice patterns in the oscillation amplitude as a function of
position along the bunch. These may be an artifact of the imperfect
synchronization between the~$A$ and $B$ signals, but they may also be
related to the physical nature of the coherent modes (i.e., rigid vs.\
soft bunch, head-on vs.\ long range). The phase of the oscillations as
a function of frequency and bunch number may also provide physical
insight. These aspects are not covered in the present analysis.

\section{Results}
\label{sec:results}

\begin{turnpage}
\begin{table*}
\caption{Summary of experimental conditions: instantaneous
  luminosity,~\Lu; average number of protons and antiprotons per
  bunch,~\Np and~\Na; average transverse emittances from the synchrotron light
monitor (95\%, normalized),~\epx, \epy, \eax, and~\eay; average longitudinal
emittances,~\epz and~\eaz\ (1 standard deviation); average incoherent
tunes from the 1.7-GHz
Schottky detector,~\qpx, \qpy, \qax, and~\qay; calculated linear
beam-beam parameters per interaction point,~\xipx, \xipy, \xiax,
and~\xiay.}
{\scriptsize
\begin{ruledtabular}
\begin{tabular}{rrrrrrrrrrrrrrrrrrrr}
Store & \multicolumn{1}{c}{Date} & \multicolumn{1}{c}{Time} & \Lu & \Np & \Na & \epx & \epy & \eax & \eay & \epz & \eaz & \qpx & \qpy & \qax & \qay & \xipx & \xipy & \xiax & \xiay \\
 & & & \q{}{10^{32}/(cm^2\,s)} & $10^{11}$ & $10^{11}$ & \q{}{\mu m} & \q{}{\mu m} & \q{}{\mu m} & \q{}{\mu m} & \q{}{eV\, s} & \q{}{eV\, s} & & & & & & & & \\
\hline
7679 & 15~Mar~2010 & 16:00 & 3.350 & 2.891 & 0.8619 & 16.9 & 23.5 &  7.9 &  8.7 & 3.29 & 3.27 & 0.6038 & 0.5914 & 0.6175 & 0.5981 & 0.0078 & 0.0074 & 0.0115 & 0.0098\\
7706 & 25~Mar~2010 & 15:13 & 3.500 & 2.953 & 0.9117 & 15.9 & 22.1 &  8.8 &  8.8 & 3.29 & 2.96 & 0.6082 & 0.6138 & 0.6235 & 0.6126 & 0.0076 & 0.0076 & 0.0125 & 0.0106\\
7706 & 25~Mar~2010 & 16:22 & 2.670 & 2.756 & 0.8144 & 17.2 & 23.7 &  8.7 &  9.7 & 3.84 & 4.03 & 0.5910 & 0.5904 & 0.5876 & 0.5860 & 0.0067 & 0.0063 & 0.0108 & 0.0092\\
7711 & 26~Mar~2010 & 12:09 & 1.430 & 2.822 & 0.5842 & 22.5 & 31.3 & 11.6 & 13.8 & 4.71 & 4.59 & 0.5888 & 0.5889 & 0.5868 & 0.5868 & 0.0035 & 0.0032 & 0.0084 & 0.0072\\
7719 &  1~Apr~2010 & 11:58 & 0.810 & 2.272 & 0.5319 & 23.8 & 35.6 & 18.3 & 18.3 & 6.33 & 5.52 & 0.5880 & 0.5881 & 0.5875 & 0.5866 & 0.0021 & 0.0021 & 0.0063 & 0.0051\\
7724 &  2~Apr~2010 & 15:38 & 2.950 & 2.791 & 0.8475 & 17.0 & 24.7 &  9.2 &  9.3 & 3.57 & 3.21 & 0.5901 & 0.5904 & 0.5908 & 0.5969 & 0.0067 & 0.0067 & 0.0109 & 0.0091\\
7724 &  2~Apr~2010 & 16:07 & 2.690 & 2.752 & 0.8197 & 17.4 & 25.4 &  9.5 &  9.8 & 3.73 & 3.39 & 0.5887 & 0.5892 & 0.5857 & 0.5907 & 0.0063 & 0.0062 & 0.0105 & 0.0087\\
7754 & 21~Apr~2010 & 16:09 & 3.510 & 2.856 & 0.9428 & 15.0 & 23.3 &  7.7 &  9.4 & 3.31 & 3.52 & 0.5978 & 0.5908 & 0.6163 & 0.6096 & 0.0085 & 0.0077 & 0.0124 & 0.0100\\
7754 & 21~Apr~2010 & 16:24 & 3.200 & 2.810 & 0.9236 & 15.1 & 23.1 &  7.9 &  9.4 & 4.41 & 3.64 & 0.5901 & 0.5899 & 0.5973 & 0.6034 & 0.0082 & 0.0075 & 0.0122 & 0.0099\\
7754 & 21~Apr~2010 & 16:45 & 2.970 & 2.769 & 0.9008 & 15.4 & 23.8 &  8.6 &  9.8 & 4.43 & 3.72 & 0.5924 & 0.5897 & 0.5864 & 0.5900 & 0.0074 & 0.0070 & 0.0117 & 0.0095\\
7754 & 21~Apr~2010 & 17:10 & 2.730 & 2.737 & 0.8806 & 15.8 & 24.4 &  9.0 & 10.0 & 4.50 & 3.85 & 0.5890 & 0.5889 & 0.5867 & 0.5872 & 0.0070 & 0.0066 & 0.0113 & 0.0091\\
7754 & 21~Apr~2010 & 21:41 & 1.570 & 2.554 & 0.7333 & 18.6 & 29.1 & 12.3 & 13.7 & 5.54 & 4.82 & 0.5883 & 0.5885 & 0.5867 & 0.5863 & 0.0043 & 0.0040 & 0.0089 & 0.0071\\
7754 & 22~Apr~2010 & 10:24 & 0.670 & 2.241 & 0.5072 & 23.8 & 36.8 & 19.6 & 20.7 & 6.85 & 6.04 & 0.5882 & 0.5887 & 0.5871 & 0.5877 & 0.0019 & 0.0018 & 0.0061 & 0.0049\\
7893 & 15~Jun~2010 & 13:26 & 0.055 & 2.703 & 0.4600 & 25.9 & 36.0 & 19.6 & 20.1 & 3.86 & 2.78 & 0.5849 & 0.5852 & 0.5639 & 0.5628 & 0.0017 & 0.0017 & 0.0070 & 0.0060\\
\end{tabular}
\end{ruledtabular}
}
\label{tab:experimental_conditions}
\end{table*}
\end{turnpage}

\begin{figure*}
\includegraphics[width=\textwidth]{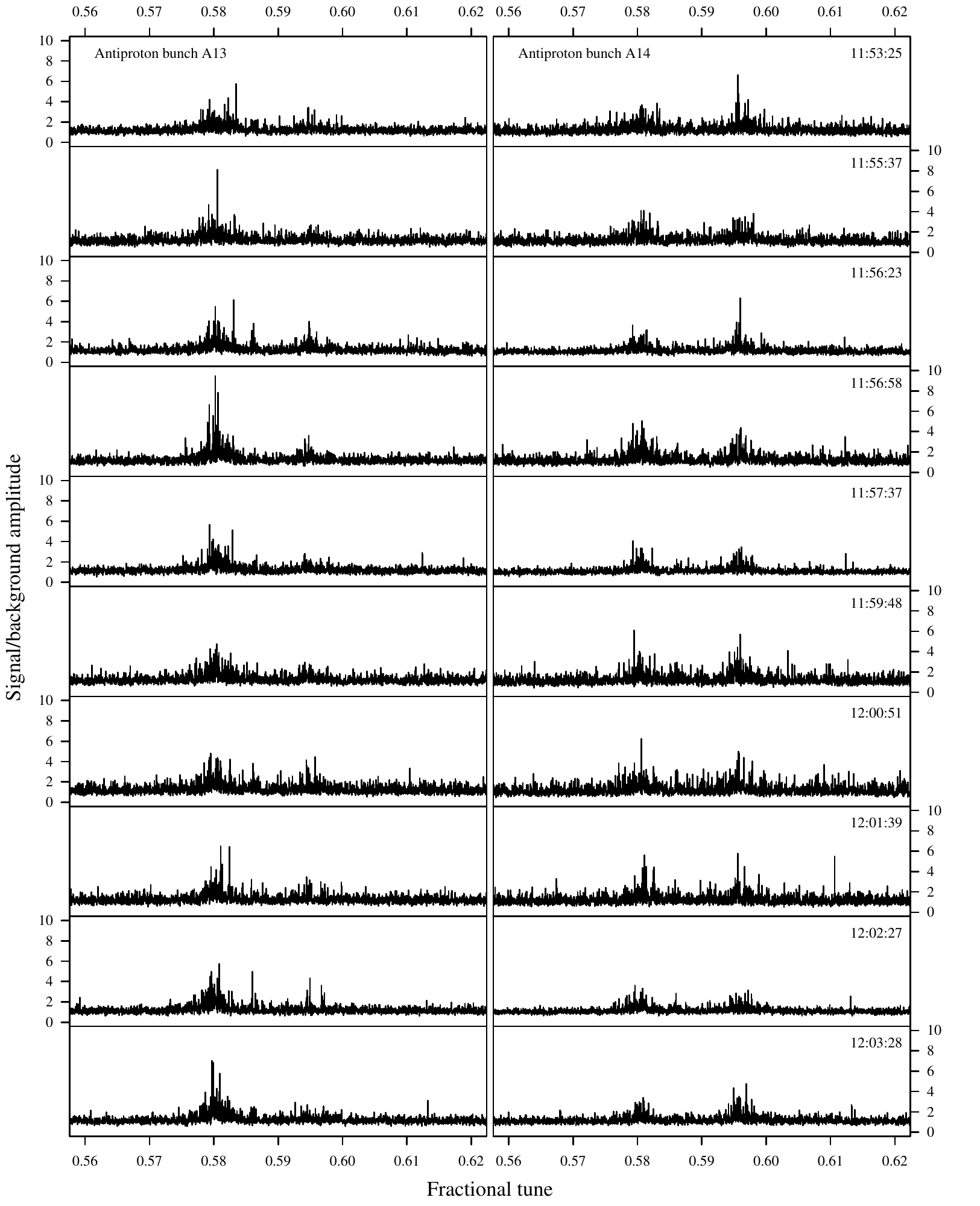}
\caption{Repeated measurements on 2 antiproton bunches during Store~7719.}
\label{fig:store7719_repeatability}
\end{figure*}

\begin{figure*}
\includegraphics[width=\textwidth]{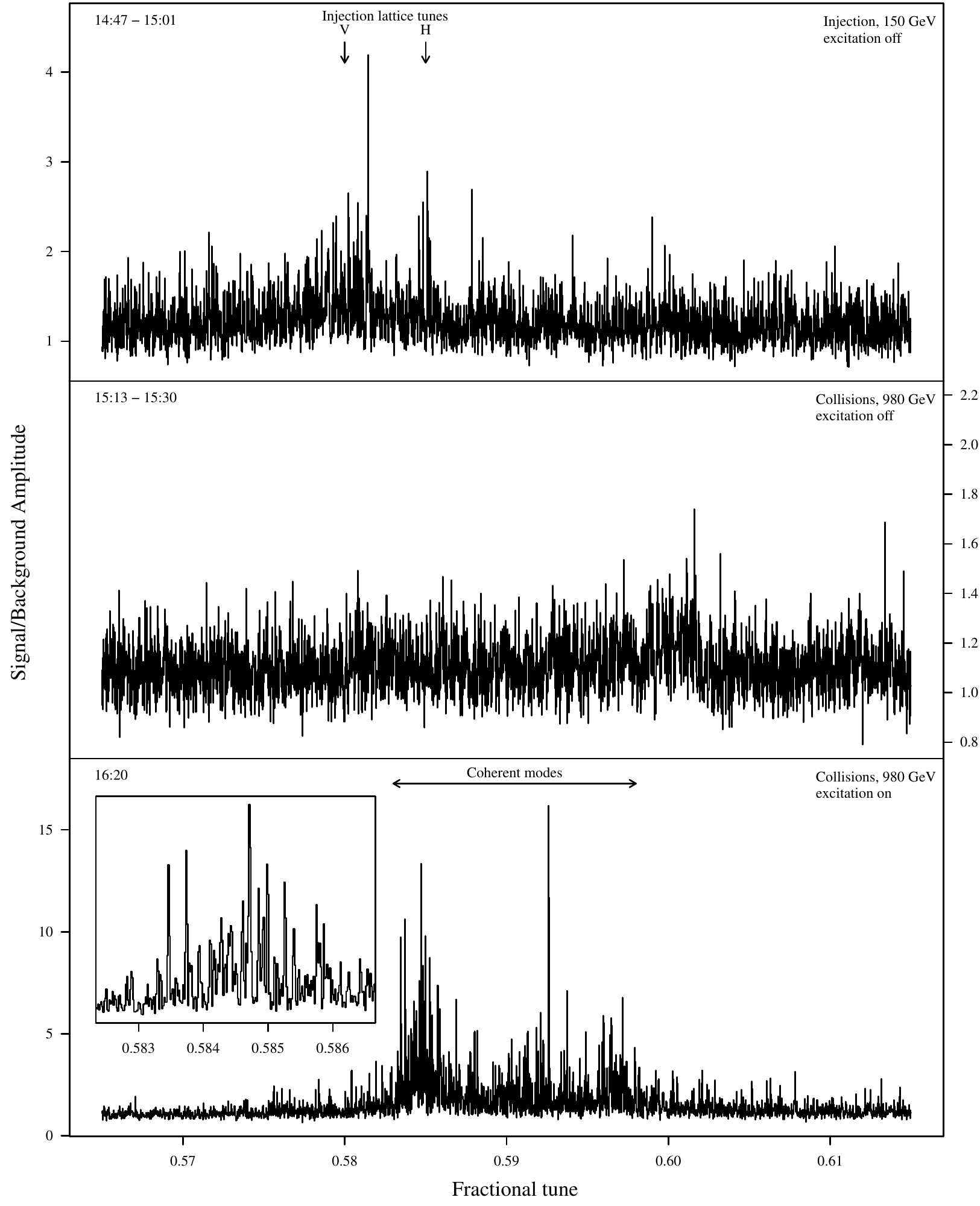}
\caption{Vertical coherent modes for one proton bunch during
  Store~7706, at injection and at collisions.}
\label{fig:store7706_comparison}
\end{figure*}

\begin{figure*}
\includegraphics[width=\textwidth]{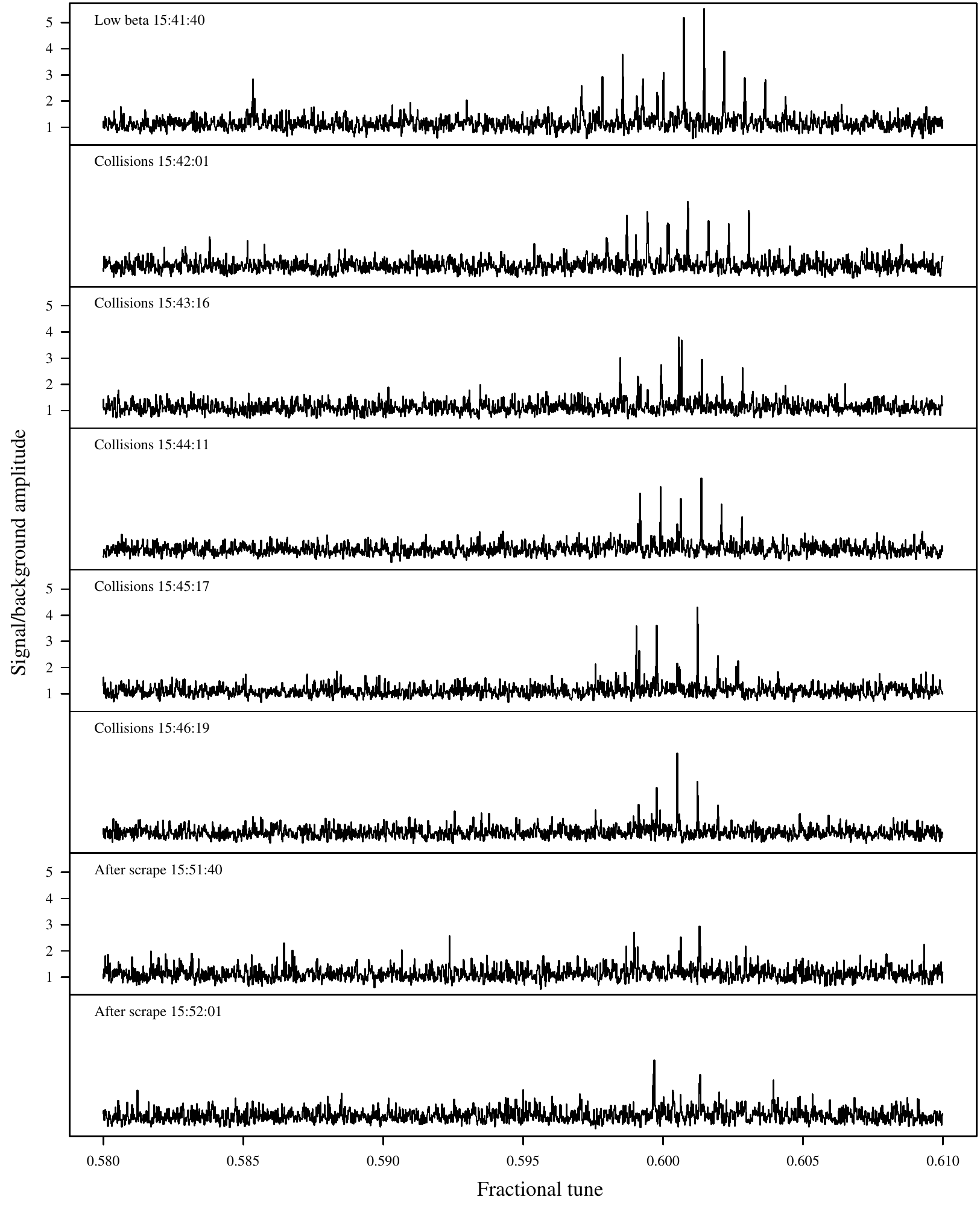}
\caption{Vertical coherent modes for antiproton bunch A25 during
  Store~7679 with no excitation, from the low-beta squeeze to after
  halo scraping.}
\label{fig:store7679_comparison}
\end{figure*}

\begin{figure*}
\includegraphics[width=\textwidth]{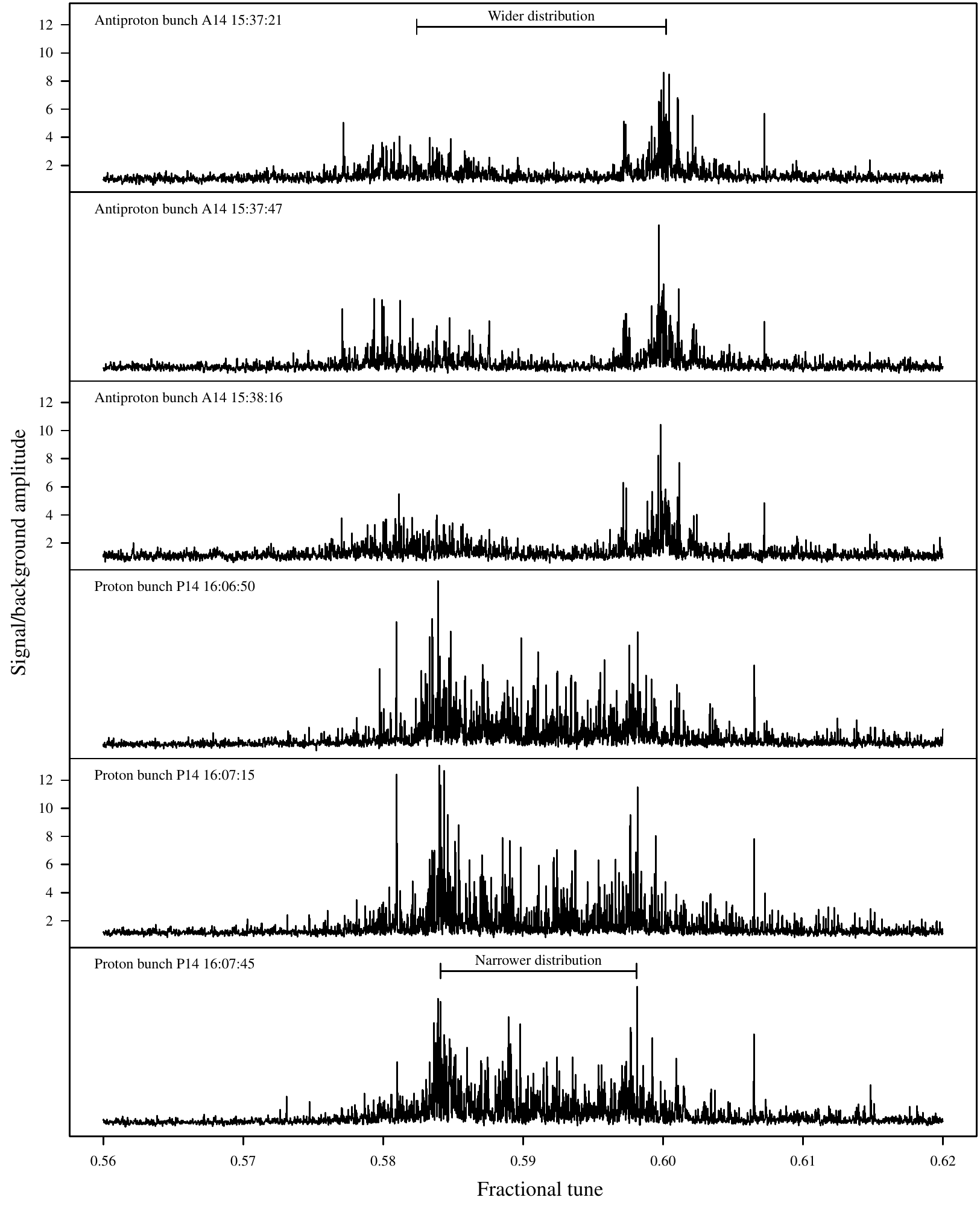}
\caption{Comparison of antiproton and proton vertical coherent modes
  during Store~7724.}
\label{fig:store7724_A14vP14}
\end{figure*}

\begin{figure*}
\includegraphics[width=\textwidth]{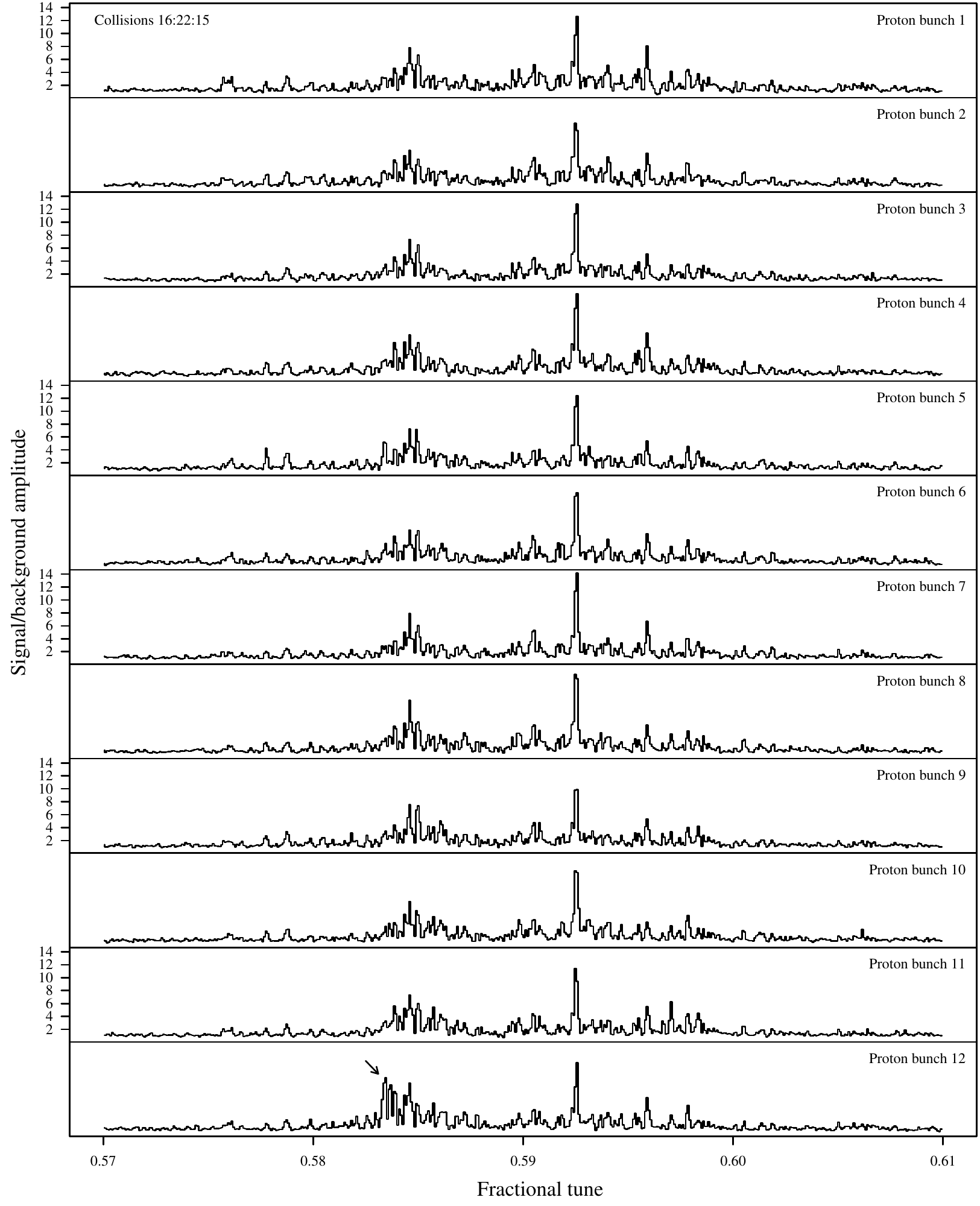}
\caption{Vertical coherent modes for all proton bunches in a train
  during Store~7706.}
\label{fig:store7706_bunch_spectra}
\end{figure*}

\begin{figure*}
\includegraphics[width=\textwidth]{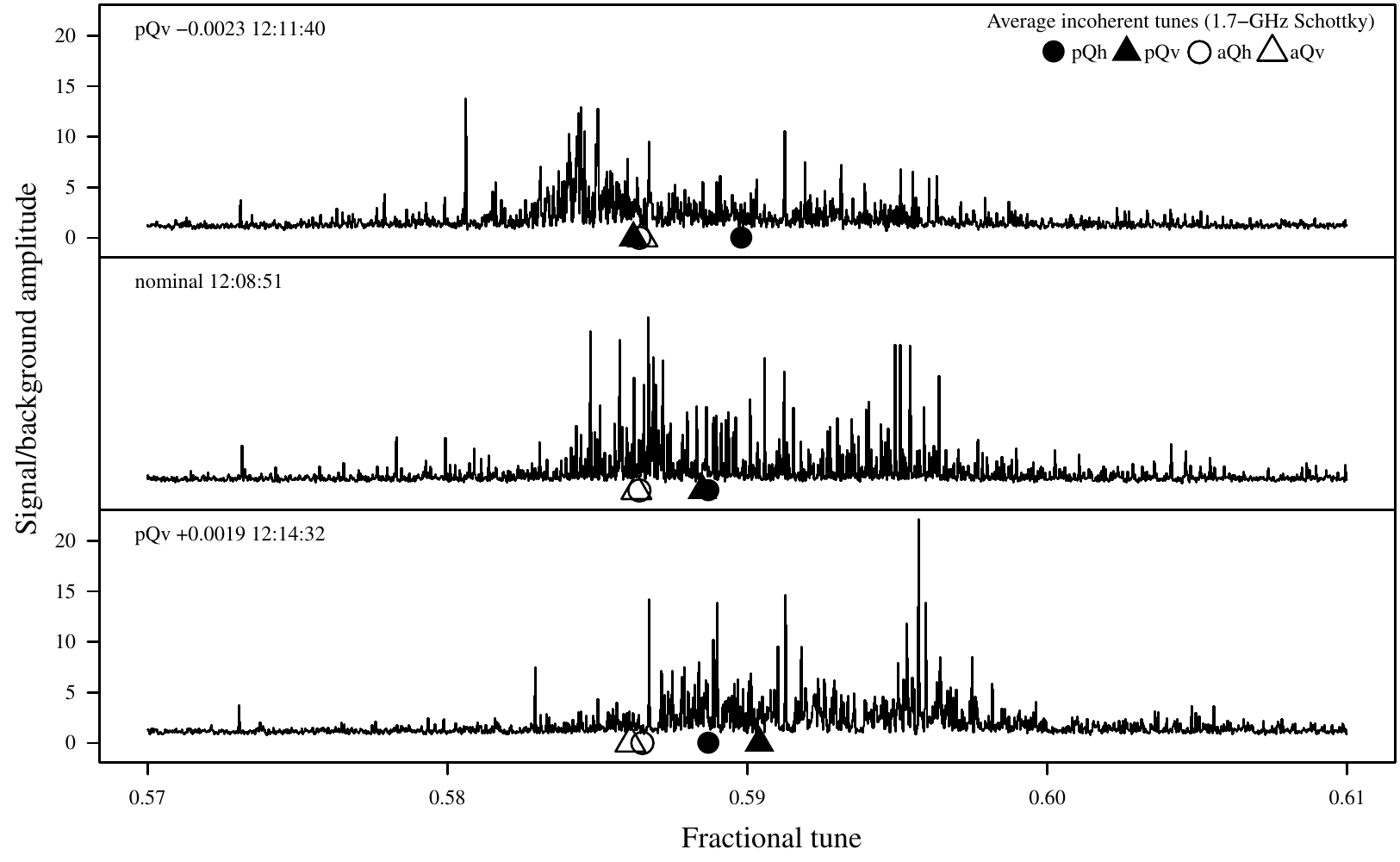}
\caption{Response of vertical coherent modes for proton bunch P11 to
  lattice tune changes during Store~7711.}
\label{fig:store7711_pQchange}
\end{figure*}

\begin{figure*}
\includegraphics[width=\textwidth]{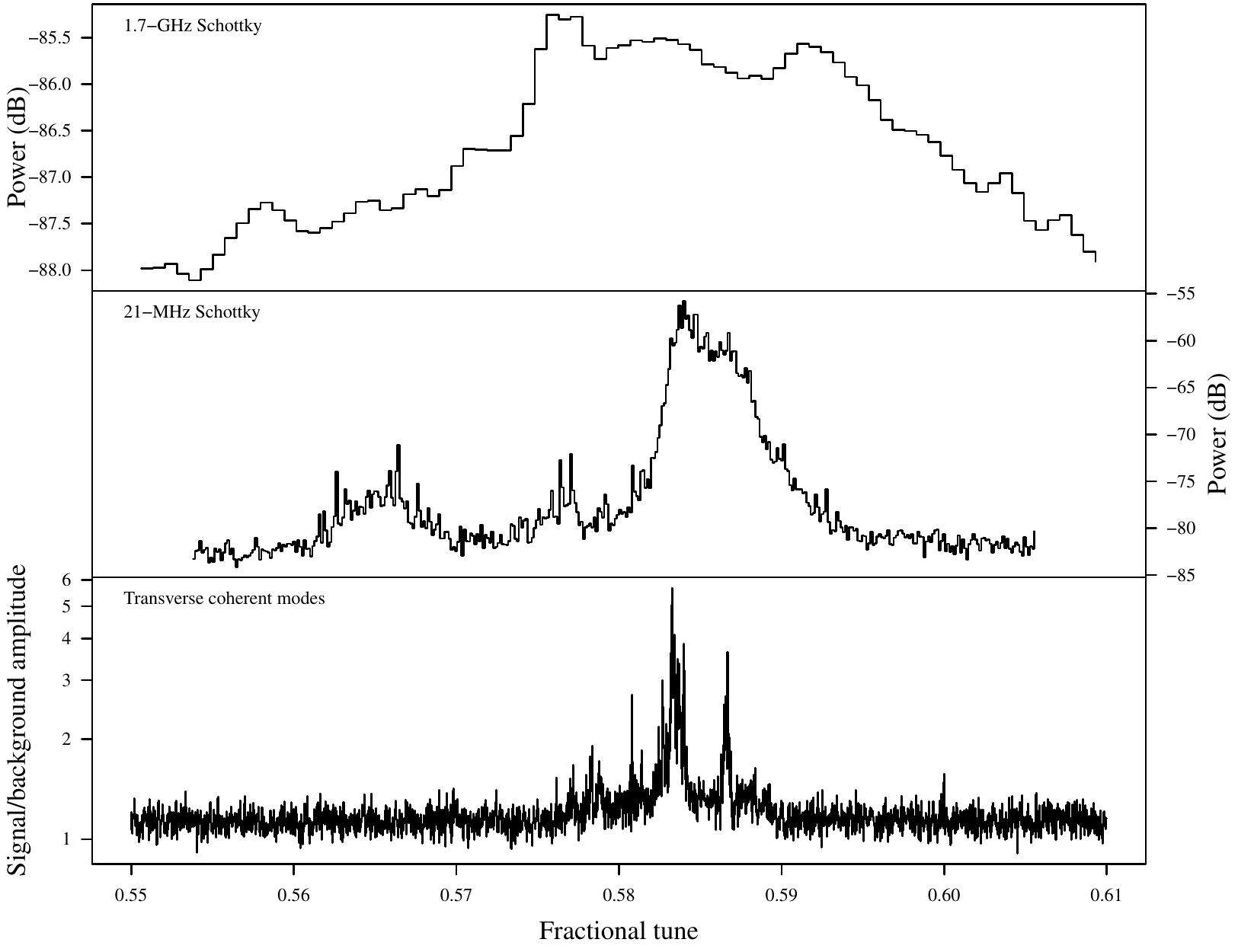}
\caption{Comparison of Schottky and coherent spectra during dedicated 3-on-3
  Store~7893.}
\label{fig:store7893_3x3_comparison}
\end{figure*}

\begin{figure*}
\includegraphics[width=\textwidth]{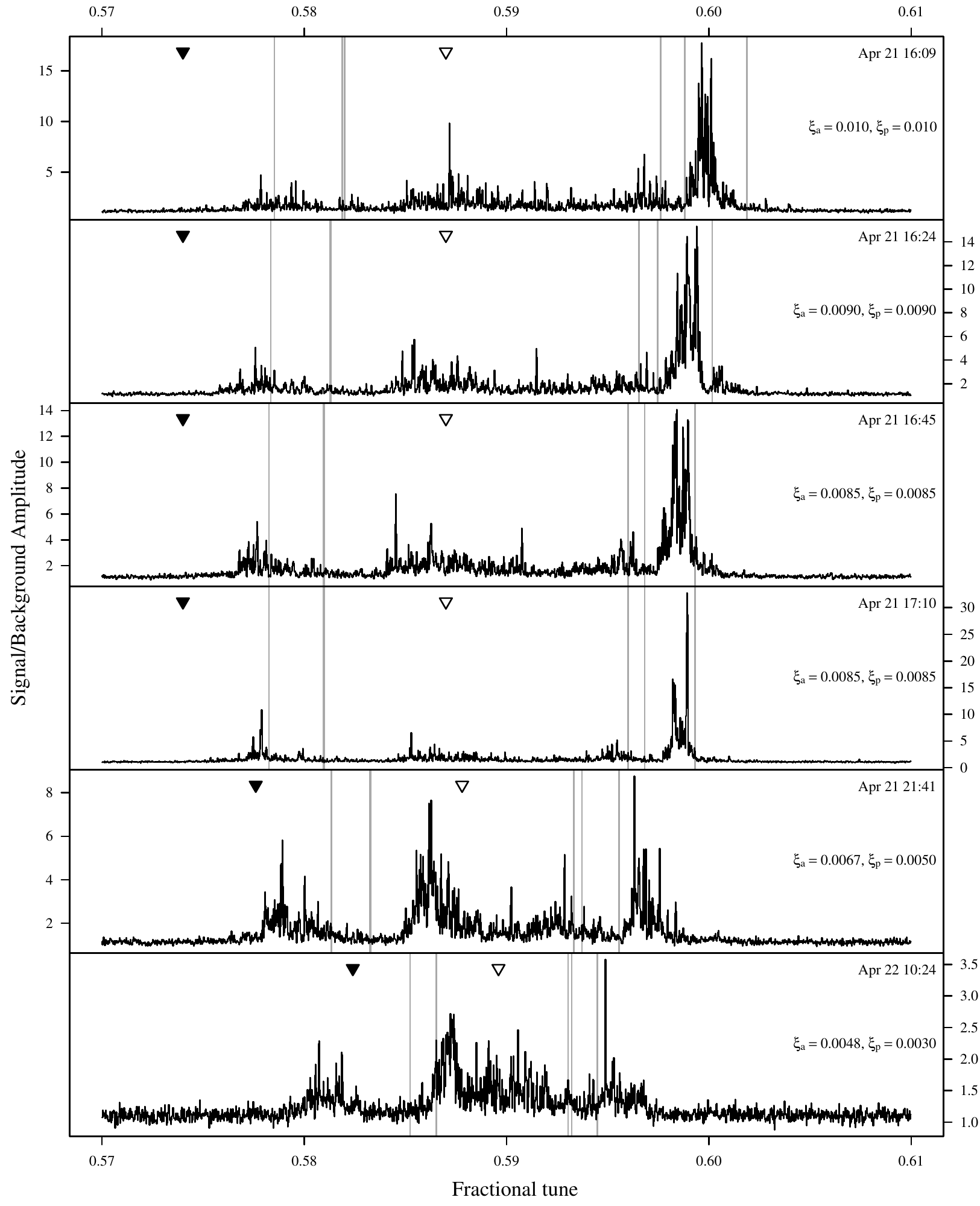}
\caption{Evolution of vertical coherent beam-beam modes for antiproton
  bunch~A13 during the course of Store~7754.}
\label{fig:store7754_evolution}
\end{figure*}

Transverse coherent mode spectra were measured for both proton and
antiproton bunches under a wide range of experimental conditions. A
few representative examples are discussed in this Section. The machine
and beam parameters relevant to the results presented below are
collected in Table~\ref{tab:experimental_conditions}.

Repeatability of the measurement was tested with antiprotons towards
the end of Store~7719, when lattice tunes were kept constant and the
beam-beam parameter was varying
slowly. Figure~\ref{fig:store7719_repeatability} shows the results of
10~consecutive measurement cycles on antiproton bunches~A13
and~A14. The frequency resolution of single modes is high (52,488
turns were analyzed in this case). One may note the stability of the
frequencies and the different distribution of amplitudes between the
two bunches. The amplitude of each mode shows some variability, as it
depends on the timing of the excitation and its duration, which were
manually controlled.

At injection (150~GeV) during Store~7706, while antiprotons were being
loaded into the machine on a separate orbit, the proton signal showed
peaks at the injection lattice tunes (0.585 in the horizontal plane
and 0.580 in the vertical one), even without excitation
(Figure~\ref{fig:store7706_comparison}, top). At collisions, it was
usually necessary to apply the excitation to see a signal. When the
excitation was applied, the signal was enhanced by at least a
factor~10 and the pattern of transverse coherent beam-beam modes
appeared (Figure~\ref{fig:store7706_comparison}, center and
bottom). The inset at the bottom of
Figure~\ref{fig:store7706_comparison} shows that the frequency
resolution is high enough to separate individual modes.

The amplitude of the signal without excitations and a comparison with
the low-frequency beam jitter allows one to estimate the absolute
magnitude of the coherent oscillations. For instance, if one takes a
typical signal-to-background ratio of 80 for the low-frequency motion,
which corresponds to an amplitude of \q{25}{\mu m}, one obtains an
amplitude of \q{60}{nm} for a signal-to-background ratio of 1.2 at the
frequencies of interest. This translates into \q{20}{nm} for the
average amplitude function of 75~m around the Tevatron ring (see also
Ref.~\cite{Shiltsev:PAC:2011}). This is to be taken as a rough
estimate, as the signal is often below the detection limit and it
varies from store to store.

Another example of signals measured without excitation is shown in
Figure~\ref{fig:store7679_comparison}. Data was collected during
Store~7679 for antiproton bunch~A25 at top energy after the low-beta
squeeze, after initiating collisions, and after scraping the halo. One
can see slowly damped oscillations around $Q=0.6$ and their
synchrotron sidebands.

During Store~7724, the signals of protons and antiprotons were
compared (Figure~\ref{fig:store7724_A14vP14}). For equal beam
intensities, emittances and tunes, one would expect to observe the
same modes in both beams. The elapsed time between the two sets of
measurements was necessary to swap cables and re-equalize the
signals. The structure of the spectra is similar, but the antiproton
distribution is wider. From this, one can infer that the antiproton
beam-beam parameter was larger. The calculated values were $\xiay =
0.009$ and $\xipy = 0.006$ at each of the 2 collision points.

An example of bunch-by-bunch measurements for protons during
Store~7706 is shown in Figure~\ref{fig:store7706_bunch_spectra}. The
signal from all 12~bunches in a train was recorded for
12,382~turns. All bunches show very similar spectra except for P12,
for which stronger lower modes are present (indicated by the arrow),
probably due to long-range interactions.

Figure~\ref{fig:store7711_pQchange} shows an example of the response
of the proton coherent mode spectra to changes in the vertical proton
lattice tune during Store~7711. The lattice setting of \qpy\ was
changed from $-0.0023$ to $+0.0019$ with respect to the nominal
value. The average incoherent tunes measured by the 1.7-GHz~Schottky
detectors are also shown for comparison (circles and triangles). In
this experiment, the vertical beam-beam parameter for protons ($\xipy
= 0.003$) was about twice as small as the one for antiprotons.

A special store with only 3 equally spaced proton bunches colliding
with 3 antiproton bunches was studied for beam-beam compensation
purposes (Store~7893). During this store, the signal from 3~systems
was recorded: the 1.7-GHz Schottky monitor, the 21-MHz Schottky
monitor, and the coherent mode detector described in this paper
(Figure~\ref{fig:store7893_3x3_comparison}). Unfortunately, transverse
beam sizes were large, making the beam-beam parameter quite
small. However, this measurement illustrates the unique features of
each system. The 1.7-GHz Schottky monitor has bunch-by-bunch
capability and good separation of the proton and antiproton
signals. It is routinely used to report the average tune of each beam
every minute. Because it operates at a large harmonic number
($h=35,631$), the sidebands contain a large number of synchrotron
satellites and their widths are dominated by the momentum spread of
the beam. The 21-MHz Schottky system has better frequency resolution,
but it cannot distinguish protons from antiprotons and it is not gated
to individual bunches. The coherent-mode detector has high frequency
resolution and fast bunch-by-bunch response. It currently requires
excitation of the beam, but this limitation can be overcome by
automatically filtering and equalizing the signals to extend its
dynamic range. The interpretation of the spectra for large beam-beam
parameters, high number of bunches, and complex collision patterns can
also be considered a limitation.

An illustration of the evolution of transverse coherent modes over a
complete store is shown in Figure~\ref{fig:store7754_evolution} for
vertical antiproton oscillations. As expected, on can see that as the
beam-beam force weakens, the spread in coherent modes decreases. Over
the course of a store, the lattice tunes need to be periodically
adjusted to keep the average incoherent tune close to the desired
working point. The bare lattice tune for antiprotons (black triangles)
and for protons (empty triangles) is estimated from the machine
settings and their calibration. The vertical gray lines represent the
prediction of the simplified model presented in
Section~\ref{sec:model} using the estimated bare lattice tunes and the
beam-beam parameters calculated from the measured beam intensities and
synchrotron-light emittances. Except for the last two measurements,
which may be affected by the evolving linear coupling and by a slight
miscalibration of the tune settings, one can see that the estimated
lattice tune lies below the first group of eigenmodes, and that the
predicted eigenfrequencies are close to the measured peaks. Obviously,
the measured spectra are richer than those predicted by the simplified
model, and a complete explanation requires a more detailed description
of the beam dynamics, such as the one found in
Ref.~\cite{Stern:PRSTAB:2010} based on a 3-dimensional strong-strong
particle-in-cell beam-beam code.

\section{Conclusions}

A system was developed to measure the spectra of coherent beam-beam
oscillations of individual bunches in the Tevatron. It is based on the
analysis of the digitized signal from a single beam-position
monitor. It requires applying band-limited noise to the beam, but an
extension of its dynamic range should be possible, if needed, so as to
operate without excitation.

The device has a response time of a few seconds, a frequency
resolution of $1.6\times 10^{-5}$ in fractional tune, and it is
sensitive to oscillation amplitudes of 60~nm. In terms of sensitivity,
resolution, and background level, it provides a very clean measurement
of coherent oscillations in hadron machines. The system complements
Schottky detectors and beam transfer function measurements as a
diagnostic tool for tunes, tune spreads, and beam-beam effects.

It was confirmed that coherent oscillations in the Tevatron are
stable, probably thanks to the different intensities of the two beams,
their tune separation, and chromaticity. The average amplitude of the
oscillations around the ring was estimated to be of the order of
20~nm.

A simplified collision model was used to calculate mode
eigenfrequencies and to show their dependence on the beam-beam
coupling. It was compared with observations made over the course of a
collider store, as the strength of the beam-beam force decreased with
time.

Spectra were acquired at different tune separations,
beam-beam parameters, and collision schemes to provide an experimental
basis for beam-beam numerical codes.

\begin{acknowledgments}
  The authors would like to thank V.~Kamerdzhiev (Forschungszentrum
  J\"ulich, Germany), F.~Emanov (Budker Institute for Nuclear Physics,
  Novosibirsk, Russia), Y.~Alexahin, B.~Fellenz, V.~Lebedev,
  G.~Saewert, V.~Scarpine, A.~Semenov, and V.~Shiltsev (Fermilab) for
  their help and insights.

  Fermilab is operated by Fermi Research Alliance, LLC under Contract
  No.~DE-AC02-07\-CH\-11359 with the United States Department of
  Energy.
\end{acknowledgments}

\onecolumngrid

\printtables

\printfigures

\end{document}